\documentclass[twocolumn,prl,showpacs,floatfix]{revtex4}
\usepackage{graphicx}
\usepackage{epsfig}
\usepackage{rotating}
\usepackage{amsmath}
\usepackage{amsfonts}
\usepackage{amssymb}
\usepackage{enumerate}
\usepackage{longtable}
\usepackage{dcolumn}% Align table columns on decimal point
\usepackage{bm}

\begin{document}

\title{ Competeing orders in spin-1 and spin-3/2 XXZ Kagome antiferromagnets: A series expansion study}

\author{J. Oitmaa }
\affiliation{School of Physics, The University of New South Wales,
Sydney 2052, Australia}

\author{R. R. P. Singh}
\affiliation{Department of Physics, University of California Davis, CA 95616, USA}

\date{\rm\today}

\begin{abstract}
We study the competition between $\sqrt{3} \times \sqrt{3}$ (RT3) and $q=0$ (Q0) magnetic orders
in spin-one and spin-$3/2$ Kagome-lattice XXZ antiferromagnets with varying XY anisotropy parameter $\Delta$,
using series expansion methods. The Hamiltonian is split into two parts: an $H_0$ which favors the
classical order in the desired pattern and an $H_1$, which is treated in perturbation theory by a series expansion.
We find that the ground state energy series for the RT3 and Q0 phases are identical up to
sixth order in the expansion, but ultimately a selection occurs, which depends on spin and the anisotropy $\Delta$. 
Results for ground state energy and the magnetization are presented.
These results are compared with recent spin-wave theory and coupled-cluster
calculations. The series results for the phase diagram are close to the predictions of spin-wave theory. 
For the spin-one model at the Heisenberg point ($\Delta=1$), our results are consistent with a vanishing order parameter,
that is an absence of a magnetically ordered phase. We also develop series expansions
for the ground state energy of the spin-one Heisenberg model in the trimerized phase. We find that the ground state energy in this
phase is lower than those of magnetically ordered ones, supporting the existence of
a spontaneously trimerized phase in this model.
\end{abstract}

%\pacs{74.70.-b,75.10.Jm,75.40.Gb,75.30.Ds}

\maketitle

Kagome lattice antiferromagnets have been studied extensively both theoretically and experimentally over the last few decades \cite{balents,klhm-e}.
There is, by now, very strong numerical evidence that the ground state of the nearest-neighbor spin-half Heisenberg model
on the Kagome-lattice is a quantum spin-liquid and has no long-range magnetic order \cite{spin-liquid}. 
However, the more general XXZ model for larger
spin and with $XY$ anisotropy may well have long-range magnetic order \cite{large-S}. Indeed, several experimental Kagome systems with large spin are known
to have magnetic long-range order \cite{kagome-lro}.  

Competing magnetic orders in these models were investigated recently by Chernyshev and Zithomirsky \cite{CZ}
using non-linear spin-wave theory, where they found a phase diagram with competing $\sqrt{3} \times \sqrt{3}$ (RT3) and $q=0$ (Q0) magnetic orders
at different spin and anisotropy $\Delta$ values. The models have also been studied recently using the coupled cluster method by Gotze and Richter\cite{GR}
who found a similar but not identical phase diagram to spin-wave theory. The main difference in the phase diagram was that in the coupled-cluster calculations
the RT3 phase occupies a singificantly bigger region of the phase diagram at the expense of the Q0 phase. The purpose of this paper is to study the competing magnetic phases by
series expansion methods. Our numerical results appear much closer to the spin-wave theory.

We consider the antiferromagnetic XXZ model on the Kagome lattice with Hamiltonian
\begin{equation}
{\cal H}= \sum_{<i,j>} (S_i^x S_j^x + \  S_i^y S_j^y + \Delta S_i^z S_j^z)
\end{equation}
where the sum is over the nearest neighbor pairs and $\Delta$ is the anisotropy parameter.
We will study spin-one and spin-$3/2$ model with various $\Delta$ values less than or equal to unity corresponding 
to XY anisotropy ($\Delta<1$)  and Heisenberg symmetry ($\Delta=1$).

To carry out the series expansion around a particular non-colinear ordered state, we rotate our
axis of quantization at each site so the local $z$ axis points along the ordering direction, and reexpress the Hamiltonian
in this rotated basis. In this basis the ferromagnetic $zz$ coupling will lead to order in the desired
classical pattern. Thus by splitting the Hamiltonian into such an Ising term and calling it
the unperturbed Hamiltonian and treating the rest of the Hamiltonian by perturbation theory,
we can calculate the properties of the system in the ordered phase \cite{book,series-reviews}. The Hamiltonian we end up
with takes the form:

\begin{equation}
{\cal H}= H_0 + \lambda H_1 + t(1-\lambda) \sum_i S_i^z,
\end{equation}
with
\begin{equation}
H_0 =-{1\over 2} \sum_{<ij>} S_i^z S_j^z
\end{equation}
and
\begin{eqnarray}
H_1=&-{1\over 8}(1+2\Delta) \sum_{<ij>} (S_i^+S_j^+ + S_i^-S_j^-) \\\nonumber
    &-{1\over 8}(1-2\Delta) \sum_{<ij>} (S_i^+S_j^- + S_i^-S_j^+) \\\nonumber
    &+{\sqrt{3}\over 4} \sum_{<ij>} \eta_{ij} \{ S_i^z(S_j^++S_j^-) - (S_i^++S_i^-)S_j^z \} \\\nonumber
\end{eqnarray}
with
\begin{eqnarray}
\eta_{ij}=&+1 \qquad  A\to B, B\to C, C\to A \\\nonumber
	  &-1 \qquad  B\to A, C\to B, A\to C \\\nonumber
\end{eqnarray}
where A,B, and C are the three sublattices.
The XXZ model of interest only arises at $\lambda=1$.
Thus, the parameter $t$ can be varied to improve convergence
as it does not play any role at $\lambda=1$. We have studied the model for
different spin and anisotropy $\Delta$. 

Our interesting finding is that regardless of spin, anisotropy $\Delta$ and redundant field value $t$, the
ground state energy for RT3 and Q0 phases are identical to $6$th order in the series expansion. This high
degree of degeneracy is reminiscent of the q-independence of the high temperature susceptibility for the classical
Kagome antiferromagnet to high orders\cite{classical-hte} and the 
high-temperature order-parameter susceptibility degeneracy for the XY pyrochlore antiferromagnets \cite{xy-pyro}.
Here, the degeneracy is for the ground state energy. The degeneracy is lifted in $7$th order. The difference between
the ground state energy for RT3 and Q0 phases are given order by order in Table I for spin-one models and in Table II for spin-$3/2$ models.

\begin{widetext}

\begin{table}[h]
\caption{Difference between ground state energy series of RT3 and Q0 phases for $S=1$ model}
\begin{tabular}{rrrrrrrr}
\hline\hline
$\ \Delta\ $ & $\ t\ $ & $n=7$ & $n=8$ & $n=9$ & $n=10$ & $n=11$ & $n=12$\\
$1.0$ & $0.0$ &  1.79281056E-05 & -.00228441779 & .00485266477 & -.0131501048 & .0244393622 & -.048490238 \\
$1.0$ & $0.5$ & -1.99373138E-05 & -.000359989266 & -4.05357305E-05 & -.000541912718 & -.000190168854 &-.000584500274 \\
$1.0$ & $1.0$ & -1.00496807E-05 & -9.61299295E-05 & -.000130394373 & -.00017789055 & -.000217321242 & -.0002516233 \\
\hline
$0.8$ & $0.0$ & .000337078044 & -.00123165186 & .0025634538 & -.00520062285 & .00822960621 & -.0138676947 \\
$0.8$ & $0.5$ & 6.06764982E-05 & -9.81439873E-05 & 4.17426967E-05 & -.000109903463 & -4.91074816E-05 & -.000138445195 \\
$0.8$ & $1.0$ & 1.58188574E-05 & -5.15123324E-06 & -1.36613326E-05 & -2.32793478E-05 & -3.57282811E-05 & -5.02830945E-05 \\
\hline
$0.6$ & $0.0$ & .000368046153 & -.000640326014 & .00145292023 & -.00238946797 & .00383705524 & -.00610563131 \\
$0.6$ & $0.5$ &  7.40771254E-05 & 4.96763165E-06 & 8.15631813E-05 & 1.48517701E-05 &  5.09012595E-05 & -5.98002479E-06 \\
$0.6$ & $1.0$ &  2.09523323E-05 & 2.48510186E-05 & 3.090978E-05 & 3.26457591E-05 & 3.16691472E-05 & 2.69258504E-05 \\
\hline
$0.4$ & $0.0$ & .00028381678  & -.000321290585  & .00085742526 & -.00125112059 &  .00222202346 & -.00350982038 \\
$0.4$ & $0.5$ &  5.95457596E-05 & 3.10517426E-05 & 8.18210584E-05 &  4.17878108E-05&  7.49216123E-05 &  3.20756294E-05 \\
$0.4$ & $1.0$ &  1.73025349E-05 & 2.68002386E-05 & 3.72787257E-05 & 4.24146522E-05 & 4.52699282E-05 & 4.47170232E-05 \\
\hline
$0.2$ & $0.0$ & .000179532034 & -.000154891221 & .000493208144 & -.00066481917 & .0012891238 & -.00201744976 \\
$0.2$ & $0.5$ &  3.88092673E-05 & 2.72091805E-05 & 6.11630232E-05 & 3.75408522E-05 & 6.30259464E-05 & 3.42523708E-05 \\
$0.2$ & $1.0$ &  1.14936167E-05 & 1.93171026E-05 & 2.83962037E-05 & 3.35527207E-05 & 3.72330216E-05 & 3.83276157E-05 \\
\hline
$0.0$ & $0.0$ &  9.51120353E-05 & -6.93197882E-05 & .000254673693 & -.000318821724 &  .000660552769  & -.0010122032 \\
$0.0$ & $0.5$ &  2.1177613E-05 & 1.64290641E-05 & 3.6276253E-05 & 2.44400256E-05 & 4.05063144E-05 & 2.45502426E-05 \\
$0.0$ & $1.0$ &  6.39053891E-06 & 1.10355978E-05 & 1.6780207E-05 & 2.0390407E-05 & 2.33497084E-05 & 2.47684216E-05 \\
\hline\hline
\end{tabular}
\end{table}

\begin{table}[h]
\caption{Difference between ground state energies of RT3 and Q0 phases for $S=3/2$ model.}
\begin{tabular}{rrrrrr}
\hline\hline
$\Delta$ & $t$ & $n=7$ & $n=8$ & $n=9$ & $n=10$ \\
$1.0$ & $0.0$ &   -.000114472141 & -.00121771703 & .00246072529  & -.00806875991 \\
$1.0$ & $0.5$ & -5.29672613E-05 & -.000398665752 & .000174231254 & -.00110052262 \\
$1.0$ & $1.0$ & -2.56833861E-05 & -.000162609093 & -9.51666575E-05 &  -.000300829022 \\
\hline
$0.8$ & $0.0$ &  .000180186675 & -.000626753577 & .0012997734 & -.00278227259 \\
$0.8$ & $0.5$ & 5.97067161E-05 & -.00012385424 & .000137564896 & -.000258061125 \\
$0.8$ & $1.0$ & 2.339967E-05 & -2.44231392E-05 & 6.79724913E-06 & -3.93092026E-05 \\
\hline
$0.6$ & $0.0$ & .000238660164 & -.000315525754  & .000746629417 & -.00111861483 \\
$0.6$ & $0.5$ &  8.41645616E-05 & -1.25172465E-05 & .000119584595 & -3.55451801E-05 \\
$0.6$ & $1.0$ &  3.46495049E-05 & 2.42397017E-05 & 4.47287003E-05 &  3.3561785E-05 \\
\hline
$0.4$ & $0.0$ &  .000197086688 & -.00015596158 & .000454414143 & -.000547182914 \\
$0.4$ & $0.5$ &  7.0831025E-05 & 2.0197586E-05 & 9.64105041E-05 & 1.41819151E-05 \\
$0.4$ & $1.0$ &  2.95715969E-05 & 3.15245715E-05 & 4.76339757E-05 & 4.39917413E-05 \\
\hline
$0.2$ & $0.0$ & .000130919478 & -7.51355574E-05 & .000273391632 & -.000286281148 \\
$0.2$ & $0.5$ & 4.76753252E-05 & 2.16031853E-05 & 6.80384984E-05 & 2.10691061E-05 \\
$0.2$ & $1.0$ & 2.00976062E-05 & 2.41489131E-05 & 3.61438392E-05 & 3.53424359E-05 \\
\hline
$0.0$ & $0.0$ &  7.27922323E-05 & -3.39489158E-05 & .000149188932 & -.000138119912 \\
$0.0$ & $0.5$ &  2.68531517E-05 & 1.42218295E-05 & 4.05437452E-05 & 1.64788846E-05 \\
$0.0$ & $1.0$ &  1.14277712E-05 & 1.43851135E-05 & 2.19416679E-05 & 2.23838979E-05 \\
\hline
\hline\hline
\end{tabular}
\end{table}

\end{widetext}

Examining Table 1 and Table II closely, it is clear that $t=0$ does not have good convergence so we
need to look at higher $t$ values. In this case all terms of the difference series become negative
for $\Delta=1$, where as all terms are positive for $\Delta \le 0.6$ for both spin-one and spin-3/2.
In other words, for $\Delta=1$ the energy is lower for the RT3 phase whereas for $\Delta\le 0.6$ the
energy is lowered for the Q0 phase. For both spin values $\Delta=0.8$ is at the boundary between
the two phases as all terms in the difference series do not have the same sign. However, adding up
all the terms shows that the energy difference is still negative for both $S=1$ and $S=3/2$.
This implies that $\Delta=0.8$ is still in the RT3 phase. This suggests a phase diagram in the
$\Delta-S$ plane which runs roughly at a constant $\Delta$ separating the two phases with a critical
$\Delta$ value a little below $0.8$. This is in remarkably good agreement with the non-linear spin
wave calculation of Chernyshev and Zhitomirsky \cite{CZ}, who find that the phase boundary occurs 
at $\Delta_c\approx 0.72$. The coupled cluster calculation of Gotze and Richter find a much larger
extent of the RT3 phase. For spin-one they find that Q0 phase exists only for $\Delta$ less than about $0.3$, while for $S=3/2$
they find that the Q0 phase only exists for $\Delta$ less than about $0.5$. Clearly the series
expansion results are much closer to the non-linear spin-wave theory. 

\begin{table}[h]
\caption{Ground state energy for $S=1$ model. The mean value
of the Pad\'e estimates for the ground state energy and the spread in the values of the different approximants are shown}
\begin{tabular}{rrrrr}
\hline\hline
Phase & $\Delta$ & $t$ & mean & spread \\
RT3 &  1.0 & 0.0 & -1.3950 &  .0013 \\
RT3 &  1.0 & 0.5 & -1.3928 &  .00014 \\
RT3 &  1.0 & 1.0 & -1.3910 & .00002 \\
Q0  &  1.0 & 0.0 & -1.3903 & .0006 \\
Q0  &  1.0 & 0.5 & -1.3890 & .0004 \\
Q0  &  1.0 & 1.0 & -1.3877 & .00012 \\
RT3 &  0.8 & 0.0 & -1.3033 & .0002 \\
RT3 &  0.8 & 0.5 & -1.3019 & .0002 \\
RT3 &  0.8 & 1.0 & -1.3012 & .0005 \\
Q0  &  0.8 & 0.0 & -1.3016 & .0001 \\
Q0  &  0.8 & 0.5 & -1.3001 & .00003 \\
Q0  &  0.8 & 1.0 & -1.2992 & .00005 \\
RT3 &  0.6 & 0.0 & -1.2215 & .0003 \\
RT3 &  0.6 & 0.5 & -1.2214 & .0003 \\
RT3 &  0.6 & 1.0 & -1.2208 & .0006 \\
Q0  &  0.6 & 0.0 & -1.2221 & .0002 \\
Q0  &  0.6 & 0.5 & -1.2213 & .0002 \\
Q0  &  0.6 & 1.0 & -1.2206 & .0002 \\
RT3 &  0.4 & 0.0 & -1.1534 & .00009 \\
RT3 &  0.4 & 0.5 & -1.1535 & .0003 \\
RT3 &  0.4 & 1.0 & -1.1531 & .00011 \\
Q0  &  0.4 & 0.0 & -1.1544 & .0002 \\
Q0  &  0.4 & 0.5 & -1.1541 & .0002 \\
Q0  &  0.4 & 1.0 & -1.1536 & .0002 \\
RT3 &  0.2 & 0.0 & -1.0987 & .00013 \\
RT3 &  0.2 & 0.5 & -1.0989 &  .00012 \\
RT3 &  0.2 & 1.0 & -1.0990 & .0003 \\
Q0  &  0.2 & 0.0 & -1.0995 & .00014 \\
Q0  &  0.2 & 0.5 & -1.0995 & .00007 \\
Q0  &  0.2 & 1.0 & -1.0996 & .00008 \\
RT3 &  0.0 & 0.0 & -1.0563 & .00003 \\
RT3 &  0.0 & 0.5 & -1.0562 & .00009 \\
RT3 &  0.0 & 1.0 & -1.0563 & .0001 \\
Q0  &  0.0 & 0.0 & -1.0568 & .00002 \\
Q0  &  0.0 & 0.5 & -1.0568 & .00008 \\
Q0  &  0.0 & 1.0 & -1.0568 & .00009 \\
\hline\hline
\end{tabular}
\end{table}

\begin{table}[h]
\caption{Ground state energy for $S=3/2$ model. The mean value
of the Pad\'e estimates for the ground state energy and the spread in the values of the different approximants are shown}
\begin{tabular}{rrrrr}
\hline\hline
Phase & $\Delta$ & $t$ & mean & spread \\
RT3 & 1.0 & 0.0 & -2.8193 & .004 \\
RT3 & 1.0 & 0.5 & -2.8250 & .004 \\
RT3 & 1.0 & 1.0 & -2.8175 & .005 \\
Q0 & 1.0 & 0.0 & -2.8185 & .004 \\
Q0 & 1.0 & 0.5 & -2.8229 & .003 \\
Q0 & 1.0 & 1.0 & -2.8162 & .004 \\
RT3 & 0.8 & 0.0 & -2.6661 & .002 \\
RT3 & 0.8 & 0.5 & -2.6674 & .002 \\
RT3 & 0.8 & 1.0 & -2.6664 & .0008 \\
Q0 & 0.8 & 0.0 & -2.6660 & .002 \\
Q0 & 0.8 & 0.5 & -2.6671 & .002 \\
Q0 & 0.8 & 1.0 & -2.6662 & .0006 \\
RT3 & 0.6 & 0.0 & -2.5481 & .0005 \\
RT3 & 0.6 & 0.5 & -2.5476 & .0007 \\
RT3 & 0.6 & 1.0 & -2.5469 & .0002 \\
Q0 & 0.6 & 0.0 & -2.5485 & .0004 \\
Q0 & 0.6 & 0.5 & -2.5481 & .0006 \\
Q0 & 0.6 & 1.0 & -2.5475 & .00005 \\
RT3 & 0.4 & 0.0 & -2.4559 & .0001 \\
RT3 & 0.4 & 0.5 & -2.4553 & .0004 \\
RT3 & 0.4 & 1.0 & -2.4548 & .00012 \\
Q0 & 0.4 & 0.0 & -2.4564 & .00011 \\
Q0 & 0.4 & 0.5 & -2.4563 & .0006 \\
Q0 & 0.4 & 1.0 & -2.4552 & .0002 \\
RT3 & 0.2 & 0.0 & -2.3830 & .00012 \\
RT3 & 0.2 & 0.5 & -2.3834 & .0006 \\
RT3 & 0.2 & 1.0 & -2.3826 & .00006 \\
Q0 & 0.2 & 0.0 & -2.3834 & .00014 \\
Q0 & 0.2 & 0.5 & -2.3838 & .0006 \\
Q0 & 0.2 & 1.0 & -2.3830 & .00009 \\
RT3 & 0.0 & 0.0 & -2.3261 & .00003 \\
RT3 & 0.0 & 0.5 & -2.3261 & .0002 \\
RT3 & 0.0 & 1.0 & -2.3261 & .00003 \\
Q0 & 0.0 & 0.0 & -2.3264 & .00004 \\
Q0 & 0.0 & 0.5 & -2.3266 & .0003 \\
Q0 & 0.0 & 1.0 & -2.3264 & .00002 \\
\hline\hline
\end{tabular}
\end{table}

The ground state energies, estimated by the use of Pad\'e approximants, are shown in Table-3 and Table-4.
In general, the energy difference between the two ordered phases is very small. The results are
consistent with simply examining the series term by term. The ground state energy is lower in the RT3
phase for $\Delta=1.0$ and $0.8$, and it is lower in the Q0 phase for $\Delta\le 0.6$.

The sublattice magnetization series is analyzed by first using a change of variables \cite{huse} to remove the
square-root singularity caused by spin-waves and then using Pad\'e approximants. Plots of the sublattice magnetization
for the phase with the lowest energy are shown in Fig.~1. For the XY model, our results of $M/S=0.86$ for spin-one and
$M/S=0.94$ for spin-$3/2$ are in excellent agreement with the results of the coupled cluster calculations \cite{GR}.
For the spin-one Heisenberg model our results suggest a vanishing sublattice magnetization or an absence of the
magnetically ordered phase.

For the spin-one Heisenberg model, several candidate ground state phases have been proposed \cite{hida}. Recent exact diagonalization
and density matrix renormalization group (DMRG) studies by Changlani and Lauchli \cite{CL} presented strong evidence for a spontaneously trimerized phase in the model.
Motivated by that study, we study the ground state energy of the trimerized phase by series expansions.

\begin{figure}
\begin{center}
 \includegraphics[width=7cm]{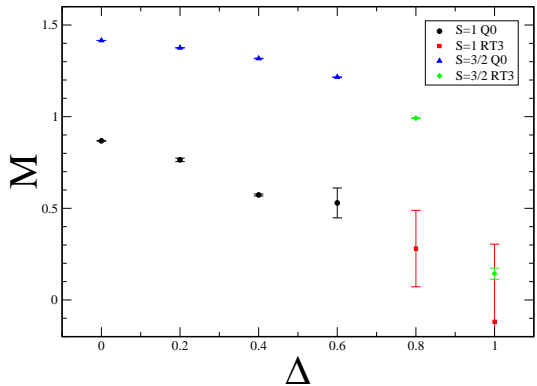}
\caption{\label{fig1}
Sublattice magnetization for the spin-one and spin-$3/2$ XXZ Kagome antiferromagnets as a function of the XY anisotropy
parameter $\Delta$. $\Delta=1$ corresponds to the Heisenberg model, where the results for the spin-one model is consistent
with a vanishing order parameter.
}
\end{center}
\end{figure}

To carry out the expansion for the trimerized phase of the spin-one model, we consider all bonds in up pointing triangles to have
exchange constant of unity, where as all bonds in down pointing triangles have exchange constant of $\alpha$. At
$\alpha=0$, this system breaks into disconnected triangles. For spin $S=1$, each triangle of spins has a unique ground state.
Series expansions can be calculated for ground state properties in powers of $\alpha$ by non-degenerate
perturbation theory \cite{book,series-reviews}.
The ground state energy per site, $e_0$, has a series expansion
\begin{eqnarray}
3e_0=& -3 - 2 \alpha^2 +{2\over 3}\alpha^3 +{11\over 18}\alpha^4 -0.33757716 \alpha^5 \\\nonumber
&-0.36266528 \alpha^6 -0.75868273\alpha^7 + \ldots \\\nonumber
\end{eqnarray}
We use Dlog Pad\'e approximants to estimate the sum of the series. The [2/4], [1/5], [3/3] and [2/3] approximants give $-4.1555$,
$-4.3801$, $-4.1391$, $-4.1236$, respectively. Upon averaging, this give an energy per site of $e_0=-1.40$, which is indeed lower than our
estimate for the energy of the ordered phases. This supports the results by Changlani and Lauchli \cite{CL} 
that the spin-one Heisenberg model has a spontaneously  trimerized ground state. 

In conclusion, in this paper we have studied the competing ground state phases of spin-one and spin-$3/2$ Kagome Lattice antiferromagnets
with XY anisotropy. We find that near the XY limit the $q=0$ magnetically ordered phase is obtained, whereas near the Heisenberg model
the $\sqrt{3} \times \sqrt{3}$ phase is realized. Our phase diagrams are in remarkably good agreement with spin-wave theory. For the spin-one
Heisenberg model, the ground state is not magnetically ordered. We presented evidence that in this case the ground state is spontaneously
trimerized.

\begin{acknowledgements}
This work is supported in part by NSF grant number  DMR-1306048, and by the computing resources provided by the Australian (APAC) National facility. 
\end{acknowledgements}

%\bibliographystyle{apsrev}
%\bibliography{../bibinput/liter10}

\end{document}